\documentclass[lengthestimate,tighten,amssymb,prl,aps,twocolumn,floatfix,tightenlines,superscriptaddress]{revtex4}
\usepackage{epsfig}
\usepackage{amsfonts}
\usepackage{amsmath}
\usepackage{epsfig}

\begin{document}
%
	\DeclareGraphicsExtensions{.eps, .jpg}

\title{Photonic crystal fibre source of photon pairs for quantum information processing}
\author{J\'er\'emie Fulconis}
\affiliation{Centre for Communications Research, Department of Electrical and Electronic Engineering, University of Bristol, Merchant Venturers Building, Woodland Road, Bristol, BS8 1UB, UK}
\author{Olivier Alibart}
\affiliation{Centre for Communications Research, Department of Electrical and Electronic Engineering, University of Bristol, Merchant Venturers Building, Woodland Road, Bristol, BS8 1UB, UK}
\author{Jeremy L. O'Brien}
\affiliation{Centre for Communications Research, Department of Electrical and Electronic Engineering, University of Bristol, Merchant Venturers Building, Woodland Road, Bristol, BS8 1UB, UK}
\affiliation{H. H. Wills Physics Laboratory, University of Bristol, Tyndall Avenue, Bristol BS8 1TL, UK}
\author{William J. Wadsworth}
\affiliation{Centre for Photonics and Photonic Materials, Department of Physics, University of Bath, Claverton Down, Bath, BA2 7AY, UK}
\author{John G. Rarity}
\affiliation{Centre for Communications Research, Department of Electrical and Electronic Engineering, University of Bristol, Merchant Venturers Building, Woodland Road, Bristol, BS8 1UB, UK}
\date{\today} 
\begin{abstract}
We demonstrate two key components for optical quantum information processing: a bright source of heralded single photons; and a bright source of entangled photon pairs.  A pair of pump photons produces a correlated pair of photons at widely spaced wavelengths (583 nm and 900 nm), via a $\chi^{(3)}$ four-wave mixing process. We demonstrate a non-classical interference between heralded photons from independent sources with a visibility of 95\%, and an entangled photon pair source, with a fidelity of 89\% with a Bell state.
\end{abstract}
\maketitle

Single photons are ideal for quantum technologies, including quantum communication \cite{gi-rmp-74-145} and quantum metrology \cite{gi-sci-306-1330}, due to intrinsically low decoherence and easy one-qubit rotations. However, realising two-qubit logic gates for quantum computation requires a massive optical nonlinearity. ÒMeasurement-inducedÓ nonlinearies can be realised using only single photon sources and detectors, and linear optical networks \cite{kn-nat-409-46}. Much progress has been made on each of these components, however, progress on linear optical networks is currently limited by the lack of bright single and pair photon sources. Here we describe a solution based on photonic crystal fibres: four wave mixing produces a correlated pair of photons at widely spaced wavelengths (583nm and 900nm). We demonstrate a bright source of heralded single photons, exhibiting a non-classical interference visibility of 95\% for independent sources; and a bright entangled pair source, with 89\% fidelity with a maximally entangled state. These sources provide an essential toolkit for photonic quantum information processing.

Since the original proposal  \cite{kn-nat-409-46} there have been a number of important theoretical improvements \cite{ni-prl-93-040503,br-prl-95-010501,ra-prl-95-100501} and experimental proof-of-principal demonstrations \cite{pi-pra-68-032316,ob-nat-426-264, ob-prl-93-080502, ga-prl-93-020504,zh-prl-94-030501,wa-nat-434-169}, which combined make optical quantum computing  promising. Experiments have typically relied on producing photons via spontaneous parametric downconversion, and detecting them with silicon avalanche photodetectors (Si APDs) with intrinsic quantum efficiencies of $\sim70\%$ and no number resolution. However, the practical limit of standard downconversion is five \cite{zh-nat-430-54,zh-prl-94-030501} or six \cite{xi-prl-97-023604} photons. Tremendous progress has been made in the development of triggered, high efficiency single photon sources  \cite{ku-prl-85-290,sa-nat-419-594,ku-nat-423-731,ku-prl-89-067901} (SPSs) and high efficiency, number resolving single photon detectors \cite{ta-apl-74-1063,ja-apl-89-031112}, however, these technolgies will not be commonplace in single photon quantum optics labs for some time. Therefore, in order to make further progress in testing single photon optical circuits for generation of large entangled cluster states, error correcting protocols, and measurement of fault tolerant thresholds,  there is an urgent practical need for a bright SPS at the visible wavelengths where Si APDs are efficient.

\begin{figure}[t!]
\begin{center}
\vspace{-0.5cm}
\includegraphics*[width=7.2cm]{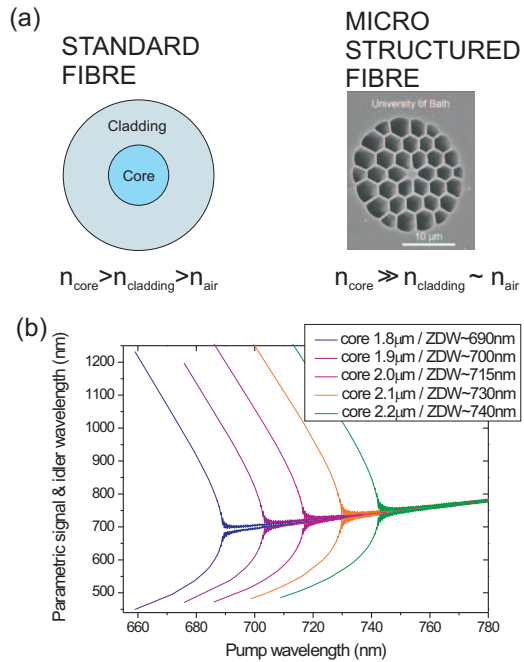}
\vspace{-0.3cm}
\caption{Photonic crystal fibres (PCFs) and four-wave mixing. (a) Conventional optical fibres are made from two different layers of silica glass: an inner core that is lightly doped to increase the refractive index; and an outer cladding of pure silica. This small difference in refractive index (typically  $\delta n\sim10^{-3}$) confines light in the core via total internal reflection. PCFs can be engineered to have much greater refractive index contrast to confine light in a very small core which leads to extremely high energy density and enhanced nonlinear effects. More importantly (for phase matching) the zero dispersion wavelength (ZDW) is shifted to shorter wavelengths as the core diameter is reduced. The scanning electron micrograph shows a PCF with a cladding consisting of 90\% air and 10\% silica, which confines light in a 2 $\mu$m core. (b) Theoretical four-wave mixing phase matching solutions for different PCF core diameters showing the shift of ZDW. By pumping the fibre below its zero dispersion wavelength in the normal dispersion regime phase matching can be realised for widely spaced wavelengths. (A more complete theoretical description of phase matching in PCFs can be found elsewhere \cite{al-njp-8-67})}
\vspace{-0.7cm}
\label{schematic}
\end{center}
\end{figure}

The development of microstructured and photonic crystal fibres \cite{ru-sci-299-358} (PCFs) enables the engineering of profoundly different optical properties to conventional optical fibres, opening the way for a range of new technologies.  PCFs with very small solid cores (Fig. \ref{schematic}(a)) can have zero dispersion wavelength (ZDW) in the visible and near infra-red region of the spectrum while the very small guided mode area leads to extremely high optical intensity giving rise to ultrahigh optical nonlinearities. Perhaps the most striking example is that of supercontinuum generation: an ultra-short pulse from a Ti-sapphire laser injected into a highly nonlinear PCF with zero dispersion close to 800 nm shows giant spectral broadening over just a few cm of fibre. Supercontinuum generation experiments primarily use wavelengths longer that the ZDW, pumping in the anamalous dispersion region. Recently pumping slightly blue shifted into the normal dispersion region has been shown to generate widely separated, phase matched, parametric amplification peaks \cite{ha-opletts-28-2225}. It is this discovery that has stimulated our recent exploration of PCF optical nonlinearities at the single photon level \cite{ra-oe-13-534,fu-oe-13-7572,fa-oe-13-5777,al-njp-8-67}.

Spontaneous parametric downconversion occurs when a pump photon spontaneously splits into two daughter photons via a three-wave mixing process in a $\chi^{(2)}$ nonlinear crystal. By carefully designing the collection scheme and pumping strategy various sources of heralded single photons and entangled photon pairs in the cw and pulsed regimes have been realised \cite{kw-prl-75-4337,ku-pra-64-023802}. However, these sources are inherently broadband and  low brightness (per nanometer, per single mode). Integrated optics has been investigated to address these issues: periodically poled waveguides of lithium niobate have been shown to be the brightest pair photon sources, but the dispersion of the medium is still an obstacle to efficient narrow-band creation and the coupling efficiency from non-circular planar waveguides into optical fibres remains poor unless significant effort is made to engineer the guided mode \cite{ta-el-37-26}. 

Until now, parametric gain from the $\chi^{(3)}$ non-linearity in optical fibres has not been seriously considered because of a lack of phase matching and an expected low efficiency compared to the $\chi^{(2)}$ in a nonlinear crystal. Even though silica glass does not possess an intrinsically high nonlinearity, the small core size in PCFs leads to strong mode confinement and high optical sensitivities even for low power (see Fig. 1). The nonlinear effects are thus amplified and give us an effective $\chi^{(3)}_{eff}$ better than $\chi^{(2)}$ in nonlinear crystals \cite{agrawal}. 

The generation of photon pairs via four-wave mixing requires two pump ($p$) photons to be converted into a signal ($s$) and idler ($i$) photon consistent with energy conservation:
\begin{equation}
\label{energy}
\frac{2}{\lambda_{p}}=\frac{1}{\lambda_{s}}+\frac{1}{\lambda_{i}}
\end{equation} 
\noindent
and phase matching:
\begin{equation}
\label{pm}
\frac{2n(\lambda_{p})}{\lambda_{p}}-2\gamma P_{p}=\frac{n(\lambda_{s})}{\lambda_{s}}+\frac{n(\lambda_{i})}{\lambda_{i}}
\end{equation} 
\noindent
where $n(\lambda_{x})$ is the refractive index at the pump, signal and idler wavelengths, $P_{p}$ is the peak pump power, and $\gamma$ is the nonlinear coefficient of the fibre:
\begin{equation}
\label{gamma}
\gamma=2\pi n_{2}/\lambda_{p}A_{eff}
\end{equation} 
\noindent
where $n_{2}=2\times 10^{-20}$ m$^{2}$/W is the nonlinear refractive index of silica (directly proportional to $\chi^{(3)}$), and $A_{eff}$ is the effective cross-sectional area of the fibre mode. 

Phase matching solutions to Eqs. \ref{energy} and \ref{pm} for different PCF core sizes at low pump power are shown in Fig. \ref{schematic}(b): by engineering the core size, the ZDW can be tuned over a wide wavelength range . In the anomalous dispersion regime the single photon pairs are close to the pump wavelength, making their separation from Raman background, inherent in guided mode configurations, problematic  \cite{li-prl-94-053601,ta-pra-72-041804, fa-oe-13-5777}. In contrast, pump wavelengths in the normal dispersion regime generate photon pairs widely spaced in wavelength  \cite{ra-oe-13-534}, making their separation from the pump, Raman background, and each other relatively straightforward \cite{ra-oe-13-534,fu-oe-13-7572, al-njp-8-67}. The exact wavelengths of the photon pair can be tuned by engineering the PCF core size (and hence ZDW), and for a given PCF can be fine tuned by adjusting the pump wavelength.

\begin{figure}[t!]
\begin{center}
\vspace{-0.4cm}
\includegraphics*[width=8.5cm]{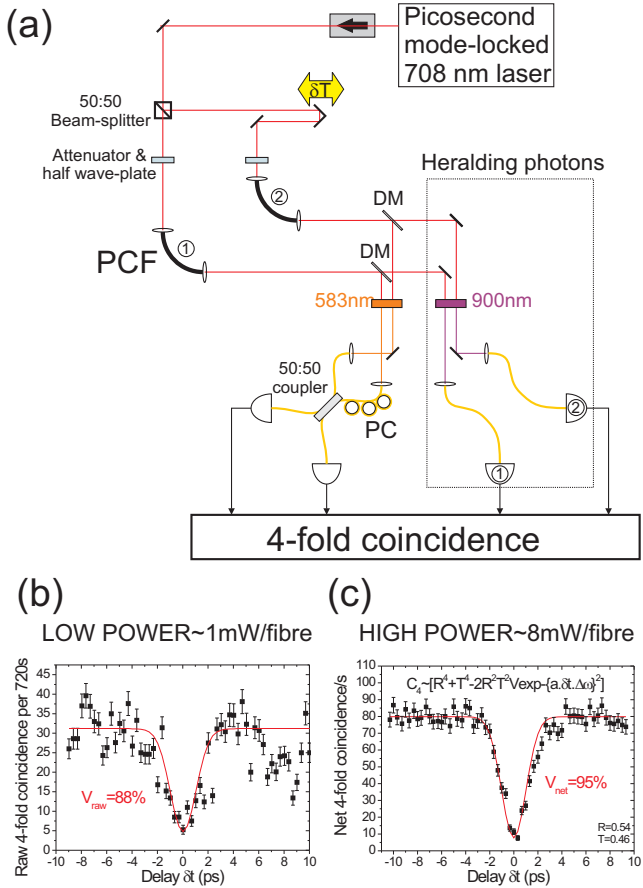}

\caption{Hong-Ou-Mandel experiment involving two heralded single photons from separate sources. (a) A picosecond pulsed laser is split in a 50:50 beamsplitter and then pumps two fibres (PCF 1 and 2, each 12 cm long) with variable delay $\delta t$. The pair photon outputs are collimated using aspheric lenses and then separated into signal and idler arms using dichroic mirrors (DM) centred at 700 nm. The signal beams are then launched into single mode fibres to guarantee optimal spatial mode overlap on the 50:50 fibre beam-splitter. A polarisation controller (PC) and 0.2 nm bandpass filter (583 nm centre wavelength) further ensured the indistinguishability of the signal photons. A slightly wider (~2 nm) filter is used on the long wavelength (900 nm) idler beams mainly to minimise losses. These beams are also collected in single mode fibre and idler detections are used to herald signal detections in a 4-fold coincidence circuit. (b) Fourfold coincidences measured as a function of delay time between the heralded photons $\delta t$ clearly showing the expected dip around zero delay. Using ~1 mW of pump power per fibre, we obtain 1700 pair coincidences/s per fibre and ~0.04 fourfold coincidences/s. (c) Increasing the pump power to 8 mW per fibre we see a net 80 fourfold coincidences per second and a high visibility dip after correction for background. }

\label{hom}
\vspace{-1cm}
\end{center}
\end{figure}

PCFs therefore can provide an extremely bright source of wavelength-tunable, polarised, narrowband single photons in a single circular spatial mode. Since photon pair creation requires annihilation of two pump photons the pair generation rate is proportional to the square of the incident intensity. A PCF source thus operates best under pulsed pumping conditions: the pair creation rate is enhanced while reducing the relative contribution from  spontaneous Raman  scattering, which is linear in pump intensity. We have previously demonstrated a PCF single photon source operating in this regime with a detected coincident photon rate of $\sim 3.2\times 10^{5}$ per second with a pump power as low as 0.5 mW \cite{fu-oe-13-7572}. In order to verify the suitability of this source for quantum information processing applications, we must first demonstrate high visibility non-classical interference between photons from different sources, and then show high fidelity production of entangled photons.     

Non-classical interference of two photons at a 50:50 beamsplitter \cite{ho-prl-59-2044} requires that the two photons be indistinguishable in every degree of freedom (wavelength, bandwidth, polarisation, spatial and temporal mode, \emph{etc.}). Figure \ref{hom}(a) is a schematic of our experiment. We demonstrate non-classical interference of heralded photons in two regimes: a low pump power regime (1 mW per fibre) where a high non-classical visibility of 88\% is observed without any need to correct for background (Fig. \ref{hom}(b)); and a higher pump power regime (8 mW per fibre) where a high non-classical visibility of 95\% is observed after subtracting the contribution from multiphoton ($>2$) contributions (Fig. \ref{hom}(c)). These visibilities are close to the theoretical maximum of 97\% based on the interference filter bandwidths. In the higher power case, we observe 80 four-fold coincidences per second,two orders of magnitude greater than previous experiments  \cite{ga-prl-93-020504}. Optimisation of the experimental setup will increase this rate and reduce the background.

\begin{figure}[t!]
\begin{center}
\vspace{-0.4cm}
\includegraphics*[width=7cm]{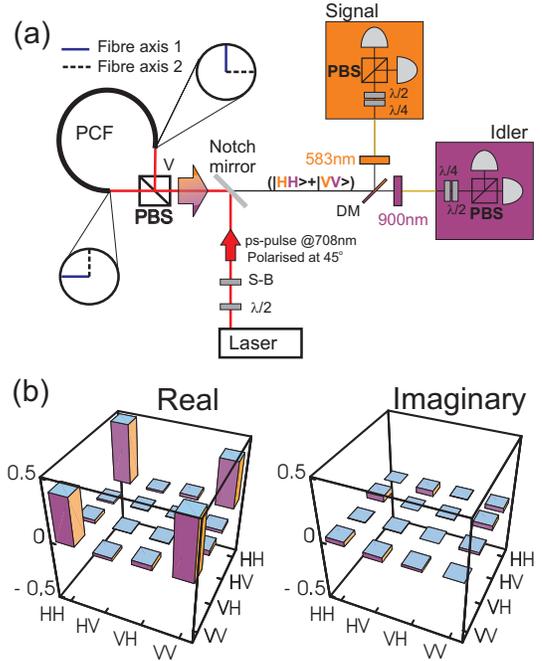}
\caption{(a) Schematic of the entangled photon pair source. The laser pulse polarisation is rotated to 45$^{\circ}$ in a half wave plate. The pump pulses are then split in a the polarizing beam-splitter (PBS) and launched into each end of the PCF. The fibre is twisted so that both pump directions are polarised along the same axis and identical phase matching conditions hold. The photon pairs thus exit from the PBS in the same port as the pumping beam and a notch mirror is used to separate them from the input pump beam. The entangled photons are launched into single mode fibre and narrow-band filtered to ensure indistinguishability between the $|HH\rangle$ and $|VV\rangle$ components and the small static phase shift between them is compensated in the pump beam by adjusting the Soleil-Babinet compensator (S-B). We use a quarter- and half-wave plate followed by a PBS to analyse the the various polarisation correlations needed to reconstruct the state \cite{ja-pra-64-052312}. (b) Tomographic reconstruction of the real and imaginary parts of the density matrix of the generated state. This clearly shows we are almost entirely in the $|HH\rangle+|VV\rangle$ state.}
\label{entangled}
\vspace{-1cm}
\end{center}
\end{figure}

For most quantum information processing applications, entangled photon pairs are an essential resource. To obtain a coherent superposition of the pair states $|HH\rangle$ and $|VV\rangle$, we use the fibre in a Sagnac interferometer with a polarizing beam-splitter as shown in Fig. \ref{entangled}. The key advantage of the Sagnac loop is that the counter-propagating pairs remain in a coherent superposition without any phase or path adjustments \cite{ta-pra-72-041804}. The pump beam polarisation is rotated to 45$^{\circ}$ so that equal intensities pump the clockwise (HH) and counter-clockwise (VV) pairs. As the fibre is slightly birefringent there are different phase matching conditions along orthogonal axes thus we twist the fibre and pump both directions along the same axis. A consequence is that the pairs exit the polarising beamsplitter in the same arm as the input pump beam and we separate them using a notch mirror.  A dichroic mirror then separates the short wavelength (583 nm) photon from the longer wavelength (900 nm). We couple both photons into single mode fibres to ensure that we only collect the fundamental modes emitted from the source.  

We use quantum state tomography and maximum likelihood techniques \cite{ja-pra-64-052312} to characterise the output from our entangled photon source. The resulting density matrix is shown in Fig. \ref{entangled}(b). This density matrix has a fidelity of $F_{\Phi^{+}}=0.89$ with the maximally entangled Bell state $|\Phi^{+}\rangle=|HH\rangle+|VV\rangle$, indicating that this is indeed a source of highly entangled photons. The main reason for the non-unit fidelity with $|\Phi^{+}\rangle$ is that the coherences between the $|HH\rangle$ and $|VV\rangle$ populations are smaller than for a pure, maximally entangled state. The degree of mixture of the state can be characterised by the linear entropy $S_{L}=0.25$, consistent with these coherences. Finally the degree of entanglement can be quantified by the tangle $T=0.63$. These results show that we have a highly entangled source of photon pairs. We suspect that the small amount of mixture observed arises from reflections at the fibre ends. We expect to improve this by optimising coincidence windows. In this experiment, with only 10 mW of pump power, the source produces $6\times 10^{3}$ coincidences per second. This figure could be improved by reducing losses in the PBS and notch mirror and we expect to reach several tens of thousands of detected entangled pairs per second with less than 20 mW pump power. 

In summary, we have described a photonic crystal fibre source of single photons for quantum information processing applications: we demonstrated that heralded single photons from independent sources undergo non-classical interference with high visibility; and demonstrated the production of highly entangled photon pairs. While this source does not meet all the requirements of the single photon sources required for scalable optical quantum computing, it represents a tremendous improvement over conventional downconversion sources: it is extremely bright, emits photon pairs which are widely tunable in wavelength, narrowband, and in a single circular spatial mode. This opens the way for more sophisticated experiments to test single photon linear optical circuits, for building large entangled clusters state, testing error correcting protocols, and directly measuring fault tolerance thresholds. These sources are also ideally suited to the next generation of linear optical circuits which will be realised in guided optics architectures, rather than bulk optics. Such devices will likely be highly wavelength dependent, making a tunable single photon source critical.

\small{\textbf{Acknowledgements}
W.J.W. is a Royal Society University Research Fellow, J.G.R. is supported by a Wolfson Merit award. The work is partly funded by UK EPSRC (QIP IRC and 1-phot), EU IP QAP and FP6-2002-IST-1-506813 SECOQC. }


\begin{thebibliography}{34}
\expandafter\ifx\csname natexlab\endcsname\relax\def\natexlab#1{#1}\fi
\expandafter\ifx\csname bibnamefont\endcsname\relax
  \def\bibnamefont#1{#1}\fi
\expandafter\ifx\csname bibfnamefont\endcsname\relax
  \def\bibfnamefont#1{#1}\fi
\expandafter\ifx\csname citenamefont\endcsname\relax
  \def\citenamefont#1{#1}\fi
\expandafter\ifx\csname url\endcsname\relax
  \def\url#1{\texttt{#1}}\fi
\expandafter\ifx\csname urlprefix\endcsname\relax\def\urlprefix{URL }\fi
\providecommand{\bibinfo}[2]{#2}
\providecommand{\eprint}[2][]{\url{#2}}

\bibitem[{\citenamefont{Gisin et~al.}(2002)\citenamefont{Gisin, Ribordy,
  Tittel, and Zbinden}}]{gi-rmp-74-145}
\bibinfo{author}{\bibfnamefont{N.}~\bibnamefont{Gisin}},
  \bibinfo{author}{\bibfnamefont{G.}~\bibnamefont{Ribordy}},
  \bibinfo{author}{\bibfnamefont{W.}~\bibnamefont{Tittel}}, \bibnamefont{and}
  \bibinfo{author}{\bibfnamefont{H.}~\bibnamefont{Zbinden}},
  \bibinfo{journal}{Rev. Mod. Phys.} \textbf{\bibinfo{volume}{74}},
  \bibinfo{pages}{145} (\bibinfo{year}{2002}).

\bibitem[{\citenamefont{Giovannetti et~al.}(2004)\citenamefont{Giovannetti,
  Lloyd, and Maccone}}]{gi-sci-306-1330}
\bibinfo{author}{\bibfnamefont{V.}~\bibnamefont{Giovannetti}},
  \bibinfo{author}{\bibfnamefont{S.}~\bibnamefont{Lloyd}}, \bibnamefont{and}
  \bibinfo{author}{\bibfnamefont{L.}~\bibnamefont{Maccone}},
  \bibinfo{journal}{Science} \textbf{\bibinfo{volume}{306}},
  \bibinfo{pages}{1330} (\bibinfo{year}{2004}).

\bibitem[{\citenamefont{Knill et~al.}(2001)\citenamefont{Knill, Laflamme, and
  Milburn}}]{kn-nat-409-46}
\bibinfo{author}{\bibfnamefont{E.}~\bibnamefont{Knill}},
  \bibinfo{author}{\bibfnamefont{R.}~\bibnamefont{Laflamme}}, \bibnamefont{and}
  \bibinfo{author}{\bibfnamefont{G.~J.} \bibnamefont{Milburn}},
  \bibinfo{journal}{Nature} \textbf{\bibinfo{volume}{409}}, \bibinfo{pages}{46}
  (\bibinfo{year}{2001}).

\bibitem[{\citenamefont{Nielsen}(2004)}]{ni-prl-93-040503}
\bibinfo{author}{\bibfnamefont{M.~A.} \bibnamefont{Nielsen}},
  \bibinfo{journal}{Phys. Rev. Lett.} \textbf{\bibinfo{volume}{93}},
  \bibinfo{eid}{040503} (\bibinfo{year}{2004}).

\bibitem[{\citenamefont{Browne and Rudolph}(2005)}]{br-prl-95-010501}
\bibinfo{author}{\bibfnamefont{D.~E.} \bibnamefont{Browne}} \bibnamefont{and}
  \bibinfo{author}{\bibfnamefont{T.}~\bibnamefont{Rudolph}},
  \bibinfo{journal}{Phys. Rev. Lett.} \textbf{\bibinfo{volume}{95}},
  \bibinfo{eid}{010501} (\bibinfo{year}{2005}).

\bibitem[{\citenamefont{Ralph et~al.}(2005)\citenamefont{Ralph, Hayes, and
  Gilchrist}}]{ra-prl-95-100501}
\bibinfo{author}{\bibfnamefont{T.~C.} \bibnamefont{Ralph}},
  \bibinfo{author}{\bibfnamefont{A.~J.~F.} \bibnamefont{Hayes}},
  \bibnamefont{and}
  \bibinfo{author}{\bibfnamefont{A.}~\bibnamefont{Gilchrist}},
  \bibinfo{journal}{Phys. Rev. Lett.} \textbf{\bibinfo{volume}{95}},
  \bibinfo{eid}{100501} (\bibinfo{year}{2005}).

\bibitem[{\citenamefont{Pittman et~al.}(2003)\citenamefont{Pittman, Fitch,
  Jacobs, and Franson}}]{pi-pra-68-032316}
\bibinfo{author}{\bibfnamefont{T.~B.} \bibnamefont{Pittman}},
  \bibinfo{author}{\bibfnamefont{M.~J.} \bibnamefont{Fitch}},
  \bibinfo{author}{\bibfnamefont{B.~C.} \bibnamefont{Jacobs}},
  \bibnamefont{and} \bibinfo{author}{\bibfnamefont{J.~D.}
  \bibnamefont{Franson}}, \bibinfo{journal}{Phys. Rev. A}
  \textbf{\bibinfo{volume}{68}}, \bibinfo{pages}{032316}
  (\bibinfo{year}{2003}).

\bibitem[{\citenamefont{O'Brien et~al.}(2003)\citenamefont{O'Brien, Pryde,
  White, Ralph, and Branning}}]{ob-nat-426-264}
\bibinfo{author}{\bibfnamefont{J.~L.} \bibnamefont{O'Brien}},
  \bibinfo{author}{\bibfnamefont{G.~J.} \bibnamefont{Pryde}},
  \bibinfo{author}{\bibfnamefont{A.~G.} \bibnamefont{White}},
  \bibinfo{author}{\bibfnamefont{T.~C.} \bibnamefont{Ralph}}, \bibnamefont{and}
  \bibinfo{author}{\bibfnamefont{D.}~\bibnamefont{Branning}},
  \bibinfo{journal}{Nature} \textbf{\bibinfo{volume}{426}},
  \bibinfo{pages}{264} (\bibinfo{year}{2003}).

\bibitem[{\citenamefont{O'Brien et~al.}(2004)\citenamefont{O'Brien, Pryde,
  Gilchrist, James, Langford, Ralph, and White}}]{ob-prl-93-080502}
\bibinfo{author}{\bibfnamefont{J.~L.} \bibnamefont{O'Brien}},
  \bibinfo{author}{\bibfnamefont{G.~J.} \bibnamefont{Pryde}},
  \bibinfo{author}{\bibfnamefont{A.}~\bibnamefont{Gilchrist}},
  \bibinfo{author}{\bibfnamefont{D.~F.~V.} \bibnamefont{James}},
  \bibinfo{author}{\bibfnamefont{N.~K.} \bibnamefont{Langford}},
  \bibinfo{author}{\bibfnamefont{T.~C.} \bibnamefont{Ralph}}, \bibnamefont{and}
  \bibinfo{author}{\bibfnamefont{A.~G.} \bibnamefont{White}},
  \bibinfo{journal}{Phys. Rev. Lett.} \textbf{\bibinfo{volume}{93}},
  \bibinfo{eid}{080502} (\bibinfo{year}{2004}).

\bibitem[{\citenamefont{Gasparoni et~al.}(2004)\citenamefont{Gasparoni, Pan,
  Walther, Rudolph, and Zeilinger}}]{ga-prl-93-020504}
\bibinfo{author}{\bibfnamefont{S.}~\bibnamefont{Gasparoni}},
  \bibinfo{author}{\bibfnamefont{J.-W.} \bibnamefont{Pan}},
  \bibinfo{author}{\bibfnamefont{P.}~\bibnamefont{Walther}},
  \bibinfo{author}{\bibfnamefont{T.}~\bibnamefont{Rudolph}}, \bibnamefont{and}
  \bibinfo{author}{\bibfnamefont{A.}~\bibnamefont{Zeilinger}},
  \bibinfo{journal}{Phys. Rev. Lett.} \textbf{\bibinfo{volume}{93}},
  \bibinfo{pages}{020504} (\bibinfo{year}{2004}).

\bibitem[{\citenamefont{Zhao et~al.}(2005)\citenamefont{Zhao, Zhang, Chen,
  Zhang, Du, Yang, and Pan}}]{zh-prl-94-030501}
\bibinfo{author}{\bibfnamefont{Z.}~\bibnamefont{Zhao}},
  \bibinfo{author}{\bibfnamefont{A.-N.} \bibnamefont{Zhang}},
  \bibinfo{author}{\bibfnamefont{Y.-A.} \bibnamefont{Chen}},
  \bibinfo{author}{\bibfnamefont{H.}~\bibnamefont{Zhang}},
  \bibinfo{author}{\bibfnamefont{J.-F.} \bibnamefont{Du}},
  \bibinfo{author}{\bibfnamefont{T.}~\bibnamefont{Yang}}, \bibnamefont{and}
  \bibinfo{author}{\bibfnamefont{J.-W.} \bibnamefont{Pan}},
  \bibinfo{journal}{Phys. Rev. Lett.} \textbf{\bibinfo{volume}{94}},
  \bibinfo{eid}{030501} (\bibinfo{year}{2005}).

\bibitem[{\citenamefont{Walther et~al.}(2005)\citenamefont{Walther, Resch,
  Rudolph, Schenck, Weinfurter, Vedral, Aspelmeyer, and
  Zeilinger}}]{wa-nat-434-169}
\bibinfo{author}{\bibfnamefont{P.}~\bibnamefont{Walther}},
  \bibinfo{author}{\bibfnamefont{K.~J.} \bibnamefont{Resch}},
  \bibinfo{author}{\bibfnamefont{T.}~\bibnamefont{Rudolph}},
  \bibinfo{author}{\bibfnamefont{E.}~\bibnamefont{Schenck}},
  \bibinfo{author}{\bibfnamefont{H.}~\bibnamefont{Weinfurter}},
  \bibinfo{author}{\bibfnamefont{V.}~\bibnamefont{Vedral}},
  \bibinfo{author}{\bibfnamefont{M.}~\bibnamefont{Aspelmeyer}},
  \bibnamefont{and}
  \bibinfo{author}{\bibfnamefont{A.}~\bibnamefont{Zeilinger}},
  \bibinfo{journal}{Nature} \textbf{\bibinfo{volume}{434}},
  \bibinfo{pages}{169} (\bibinfo{year}{2005}).

\bibitem[{\citenamefont{Zhao et~al.}(2004)\citenamefont{Zhao, Chen, Zhang,
  Yang, Briegel, and Pan}}]{zh-nat-430-54}
\bibinfo{author}{\bibfnamefont{Z.}~\bibnamefont{Zhao}},
  \bibinfo{author}{\bibfnamefont{Y.-A.} \bibnamefont{Chen}},
  \bibinfo{author}{\bibfnamefont{A.-N.} \bibnamefont{Zhang}},
  \bibinfo{author}{\bibfnamefont{T.}~\bibnamefont{Yang}},
  \bibinfo{author}{\bibfnamefont{H.~J.} \bibnamefont{Briegel}},
  \bibnamefont{and} \bibinfo{author}{\bibfnamefont{J.-W.} \bibnamefont{Pan}},
  \bibinfo{journal}{Nature} \textbf{\bibinfo{volume}{430}}, \bibinfo{pages}{54}
  (\bibinfo{year}{2004}).

\bibitem[{\citenamefont{Xiang et~al.}(2006)\citenamefont{Xiang, Huang, Sun,
  Zhang, Ou, and Guo}}]{xi-prl-97-023604}
\bibinfo{author}{\bibfnamefont{G.~Y.} \bibnamefont{Xiang}},
  \bibinfo{author}{\bibfnamefont{Y.~F.} \bibnamefont{Huang}},
  \bibinfo{author}{\bibfnamefont{F.~W.} \bibnamefont{Sun}},
  \bibinfo{author}{\bibfnamefont{P.}~\bibnamefont{Zhang}},
  \bibinfo{author}{\bibfnamefont{Z.~Y.} \bibnamefont{Ou}}, \bibnamefont{and}
  \bibinfo{author}{\bibfnamefont{G.~C.} \bibnamefont{Guo}},
  \bibinfo{journal}{Phys. Rev. Lett.} \textbf{\bibinfo{volume}{97}},
  \bibinfo{eid}{023604} (\bibinfo{year}{2006}).

\bibitem[{\citenamefont{Kurtsiefer et~al.}(2000)\citenamefont{Kurtsiefer,
  Mayer, Zarda, and Weinfurter}}]{ku-prl-85-290}
\bibinfo{author}{\bibfnamefont{C.}~\bibnamefont{Kurtsiefer}},
  \bibinfo{author}{\bibfnamefont{S.}~\bibnamefont{Mayer}},
  \bibinfo{author}{\bibfnamefont{P.}~\bibnamefont{Zarda}}, \bibnamefont{and}
  \bibinfo{author}{\bibfnamefont{H.}~\bibnamefont{Weinfurter}},
  \bibinfo{journal}{Phys. Rev. Lett.} \textbf{\bibinfo{volume}{85}},
  \bibinfo{pages}{290} (\bibinfo{year}{2000}).

\bibitem[{\citenamefont{Santori et~al.}(2002)\citenamefont{Santori, Fattal,
  Vu{\"{c}}kovi{\"{c}}, Solomon, and Yamamoto}}]{sa-nat-419-594}
\bibinfo{author}{\bibfnamefont{C.}~\bibnamefont{Santori}},
  \bibinfo{author}{\bibfnamefont{D.}~\bibnamefont{Fattal}},
  \bibinfo{author}{\bibfnamefont{J.}~\bibnamefont{Vu{\"{c}}kovi{\"{c}}}},
  \bibinfo{author}{\bibfnamefont{G.~S.} \bibnamefont{Solomon}},
  \bibnamefont{and} \bibinfo{author}{\bibfnamefont{Y.}~\bibnamefont{Yamamoto}},
  \bibinfo{journal}{Nature} \textbf{\bibinfo{volume}{419}},
  \bibinfo{pages}{594} (\bibinfo{year}{2002}).

\bibitem[{\citenamefont{Kuzmich et~al.}(2003)\citenamefont{Kuzmich, Bowen,
  Boozer, Boca, Chou, Duan, and Kimble}}]{ku-nat-423-731}
\bibinfo{author}{\bibfnamefont{A.}~\bibnamefont{Kuzmich}},
  \bibinfo{author}{\bibfnamefont{W.~P.} \bibnamefont{Bowen}},
  \bibinfo{author}{\bibfnamefont{A.~D.} \bibnamefont{Boozer}},
  \bibinfo{author}{\bibfnamefont{A.}~\bibnamefont{Boca}},
  \bibinfo{author}{\bibfnamefont{C.~W.} \bibnamefont{Chou}},
  \bibinfo{author}{\bibfnamefont{L.}~\bibnamefont{Duan}}, \bibnamefont{and}
  \bibinfo{author}{\bibfnamefont{H.~J.} \bibnamefont{Kimble}},
  \bibinfo{journal}{Nature} \textbf{\bibinfo{volume}{423}},
  \bibinfo{pages}{731} (\bibinfo{year}{2003}).

\bibitem[{\citenamefont{Kuhn et~al.}(2002)\citenamefont{Kuhn, Hennrich, and
  Rempe}}]{ku-prl-89-067901}
\bibinfo{author}{\bibfnamefont{A.}~\bibnamefont{Kuhn}},
  \bibinfo{author}{\bibfnamefont{M.}~\bibnamefont{Hennrich}}, \bibnamefont{and}
  \bibinfo{author}{\bibfnamefont{G.}~\bibnamefont{Rempe}},
  \bibinfo{journal}{Phys. Rev. Lett.} \textbf{\bibinfo{volume}{89}},
  \bibinfo{pages}{067901} (\bibinfo{year}{2002}).

\bibitem[{\citenamefont{Takeuchi et~al.}(1999)\citenamefont{Takeuchi, Kim,
  Yamamoto, and Hogue}}]{ta-apl-74-1063}
\bibinfo{author}{\bibfnamefont{S.}~\bibnamefont{Takeuchi}},
  \bibinfo{author}{\bibfnamefont{J.}~\bibnamefont{Kim}},
  \bibinfo{author}{\bibfnamefont{Y.}~\bibnamefont{Yamamoto}}, \bibnamefont{and}
  \bibinfo{author}{\bibfnamefont{H.~H.} \bibnamefont{Hogue}},
  \bibinfo{journal}{Appl. Phys. Lett.} \textbf{\bibinfo{volume}{74}},
  \bibinfo{pages}{1063} (\bibinfo{year}{1999}).

\bibitem[{\citenamefont{Jaspan et~al.}(2006)\citenamefont{Jaspan, Habif,
  Hadfield, and Nam}}]{ja-apl-89-031112}
\bibinfo{author}{\bibfnamefont{M.~A.} \bibnamefont{Jaspan}},
  \bibinfo{author}{\bibfnamefont{J.~L.} \bibnamefont{Habif}},
  \bibinfo{author}{\bibfnamefont{R.~H.} \bibnamefont{Hadfield}},
  \bibnamefont{and} \bibinfo{author}{\bibfnamefont{S.~W.} \bibnamefont{Nam}},
  \textbf{\bibinfo{volume}{89}}, \bibinfo{eid}{031112} (\bibinfo{year}{2006}).

\bibitem[{\citenamefont{Alibart et~al.}(2006)\citenamefont{Alibart, Fulconis,
  Wong, Murdoch, Wadsworth, and Rarity}}]{al-njp-8-67}
\bibinfo{author}{\bibfnamefont{O.}~\bibnamefont{Alibart}},
  \bibinfo{author}{\bibfnamefont{J.}~\bibnamefont{Fulconis}},
  \bibinfo{author}{\bibfnamefont{G.~K.~L.} \bibnamefont{Wong}},
  \bibinfo{author}{\bibfnamefont{S.~G.} \bibnamefont{Murdoch}},
  \bibinfo{author}{\bibfnamefont{W.~J.} \bibnamefont{Wadsworth}},
  \bibnamefont{and} \bibinfo{author}{\bibfnamefont{J.~G.}
  \bibnamefont{Rarity}}, \bibinfo{journal}{N. J. Phys}
  \textbf{\bibinfo{volume}{8}}, \bibinfo{pages}{67} (\bibinfo{year}{2006}).

\bibitem[{\citenamefont{Russell}(2003)}]{ru-sci-299-358}
\bibinfo{author}{\bibfnamefont{P.}~\bibnamefont{Russell}},
  \bibinfo{journal}{Science} \textbf{\bibinfo{volume}{299}},
  \bibinfo{pages}{358} (\bibinfo{year}{2003}).

\bibitem[{\citenamefont{Harvey et~al.}(2003)\citenamefont{Harvey, Leonhardt,
  Coen, Wong, Knight, Wadsworth, and Russell}}]{ha-opletts-28-2225}
\bibinfo{author}{\bibfnamefont{J.}~\bibnamefont{Harvey}},
  \bibinfo{author}{\bibfnamefont{R.}~\bibnamefont{Leonhardt}},
  \bibinfo{author}{\bibfnamefont{S.}~\bibnamefont{Coen}},
  \bibinfo{author}{\bibfnamefont{G.}~\bibnamefont{Wong}},
  \bibinfo{author}{\bibfnamefont{J.}~\bibnamefont{Knight}},
  \bibinfo{author}{\bibfnamefont{W.}~\bibnamefont{Wadsworth}},
  \bibnamefont{and} \bibinfo{author}{\bibfnamefont{P.}~\bibnamefont{Russell}},
  \bibinfo{journal}{Opt. Lett.} \textbf{\bibinfo{volume}{28}},
  \bibinfo{pages}{2225} (\bibinfo{year}{2003}).

\bibitem[{\citenamefont{Rarity et~al.}(2005)\citenamefont{Rarity, Fulconis,
  Duligall, Wadsworth, and Russell}}]{ra-oe-13-534}
\bibinfo{author}{\bibfnamefont{J.~G.} \bibnamefont{Rarity}},
  \bibinfo{author}{\bibfnamefont{J.}~\bibnamefont{Fulconis}},
  \bibinfo{author}{\bibfnamefont{J.}~\bibnamefont{Duligall}},
  \bibinfo{author}{\bibfnamefont{W.~J.} \bibnamefont{Wadsworth}},
  \bibnamefont{and} \bibinfo{author}{\bibfnamefont{P.~S.~J.}
  \bibnamefont{Russell}}, \bibinfo{journal}{Opt. Express}
  \textbf{\bibinfo{volume}{13}}, \bibinfo{pages}{534} (\bibinfo{year}{2005}).

\bibitem[{\citenamefont{Fulconis et~al.}(2005)\citenamefont{Fulconis, Alibart,
  Wadsworth, Russell, and Rarity}}]{fu-oe-13-7572}
\bibinfo{author}{\bibfnamefont{J.}~\bibnamefont{Fulconis}},
  \bibinfo{author}{\bibfnamefont{O.}~\bibnamefont{Alibart}},
  \bibinfo{author}{\bibfnamefont{W.~J.} \bibnamefont{Wadsworth}},
  \bibinfo{author}{\bibfnamefont{P.~S.} \bibnamefont{Russell}},
  \bibnamefont{and} \bibinfo{author}{\bibfnamefont{J.~G.}
  \bibnamefont{Rarity}}, \bibinfo{journal}{Opt. Express}
  \textbf{\bibinfo{volume}{13}}, \bibinfo{pages}{7572} (\bibinfo{year}{2005}).

\bibitem[{\citenamefont{Fan and Migdall}(2005)}]{fa-oe-13-5777}
\bibinfo{author}{\bibfnamefont{J.}~\bibnamefont{Fan}} \bibnamefont{and}
  \bibinfo{author}{\bibfnamefont{A.}~\bibnamefont{Migdall}},
  \bibinfo{journal}{Opt. Express} \textbf{\bibinfo{volume}{13}},
  \bibinfo{pages}{5777} (\bibinfo{year}{2005}).

\bibitem[{\citenamefont{Kwiat et~al.}(1995)\citenamefont{Kwiat, Mattle,
  Weinfurter, Zeilinger, Sergienko, and Shih}}]{kw-prl-75-4337}
\bibinfo{author}{\bibfnamefont{P.~G.} \bibnamefont{Kwiat}},
  \bibinfo{author}{\bibfnamefont{K.}~\bibnamefont{Mattle}},
  \bibinfo{author}{\bibfnamefont{H.}~\bibnamefont{Weinfurter}},
  \bibinfo{author}{\bibfnamefont{A.}~\bibnamefont{Zeilinger}},
  \bibinfo{author}{\bibfnamefont{A.~V.} \bibnamefont{Sergienko}},
  \bibnamefont{and} \bibinfo{author}{\bibfnamefont{Y.}~\bibnamefont{Shih}},
  \bibinfo{journal}{Phys. Rev. Lett.} \textbf{\bibinfo{volume}{75}},
  \bibinfo{pages}{4337} (\bibinfo{year}{1995}).

\bibitem[{\citenamefont{Kurtsiefer et~al.}(2001)\citenamefont{Kurtsiefer,
  Oberparleiter, and Weinfurter}}]{ku-pra-64-023802}
\bibinfo{author}{\bibfnamefont{C.}~\bibnamefont{Kurtsiefer}},
  \bibinfo{author}{\bibfnamefont{M.}~\bibnamefont{Oberparleiter}},
  \bibnamefont{and}
  \bibinfo{author}{\bibfnamefont{H.}~\bibnamefont{Weinfurter}},
  \bibinfo{journal}{Phys. Rev. A} \textbf{\bibinfo{volume}{64}},
  \bibinfo{pages}{023802} (\bibinfo{year}{2001}).

\bibitem[{\citenamefont{Tanzilli et~al.}(2001)\citenamefont{Tanzilli,
  de~Riedmatten, Tittel, Zbinden, Baldi, de~Micheli, Ostrowski, and
  Gisin}}]{ta-el-37-26}
\bibinfo{author}{\bibfnamefont{S.}~\bibnamefont{Tanzilli}},
  \bibinfo{author}{\bibfnamefont{H.}~\bibnamefont{de~Riedmatten}},
  \bibinfo{author}{\bibfnamefont{W.}~\bibnamefont{Tittel}},
  \bibinfo{author}{\bibfnamefont{H.}~\bibnamefont{Zbinden}},
  \bibinfo{author}{\bibfnamefont{P.}~\bibnamefont{Baldi}},
  \bibinfo{author}{\bibfnamefont{M.}~\bibnamefont{de~Micheli}},
  \bibinfo{author}{\bibfnamefont{D.~B.} \bibnamefont{Ostrowski}},
  \bibnamefont{and} \bibinfo{author}{\bibfnamefont{N.}~\bibnamefont{Gisin}},
  \bibinfo{journal}{Electron. Lett.} \textbf{\bibinfo{volume}{37}},
  \bibinfo{pages}{26} (\bibinfo{year}{2001}).

\bibitem[{\citenamefont{Agrawal}(1995)}]{agrawal}
\bibinfo{author}{\bibfnamefont{G.}~\bibnamefont{Agrawal}},
  \emph{\bibinfo{title}{Nonlinear fiber optics}} (\bibinfo{publisher}{Academic
  Press}, \bibinfo{year}{1995}), \bibinfo{note}{p 91}.

\bibitem[{\citenamefont{Li et~al.}(2005)\citenamefont{Li, Voss, Sharping, and
  Kumar}}]{li-prl-94-053601}
\bibinfo{author}{\bibfnamefont{X.}~\bibnamefont{Li}},
  \bibinfo{author}{\bibfnamefont{P.~L.} \bibnamefont{Voss}},
  \bibinfo{author}{\bibfnamefont{J.~E.} \bibnamefont{Sharping}},
  \bibnamefont{and} \bibinfo{author}{\bibfnamefont{P.}~\bibnamefont{Kumar}},
  \bibinfo{journal}{Phys. Rev. Lett.} \textbf{\bibinfo{volume}{94}},
  \bibinfo{eid}{053601} (\bibinfo{year}{2005}).

\bibitem[{\citenamefont{Takesue and Inoue}(2005)}]{ta-pra-72-041804}
\bibinfo{author}{\bibfnamefont{H.}~\bibnamefont{Takesue}} \bibnamefont{and}
  \bibinfo{author}{\bibfnamefont{K.}~\bibnamefont{Inoue}},
  \bibinfo{journal}{Phys. Rev. A} \textbf{\bibinfo{volume}{72}},
  \bibinfo{eid}{041804} (\bibinfo{year}{2005}).

\bibitem[{\citenamefont{Hong et~al.}(1987)\citenamefont{Hong, Ou, and
  Mandel}}]{ho-prl-59-2044}
\bibinfo{author}{\bibfnamefont{C.~K.} \bibnamefont{Hong}},
  \bibinfo{author}{\bibfnamefont{Z.~Y.} \bibnamefont{Ou}}, \bibnamefont{and}
  \bibinfo{author}{\bibfnamefont{L.}~\bibnamefont{Mandel}},
  \bibinfo{journal}{Phys. Rev. Lett.} \textbf{\bibinfo{volume}{59}},
  \bibinfo{pages}{2044} (\bibinfo{year}{1987}).

\bibitem[{\citenamefont{James et~al.}(2001)\citenamefont{James, Kwiat, Munro,
  and White}}]{ja-pra-64-052312}
\bibinfo{author}{\bibfnamefont{D.~F.~V.} \bibnamefont{James}},
  \bibinfo{author}{\bibfnamefont{P.~G.} \bibnamefont{Kwiat}},
  \bibinfo{author}{\bibfnamefont{W.~J.} \bibnamefont{Munro}}, \bibnamefont{and}
  \bibinfo{author}{\bibfnamefont{A.~G.} \bibnamefont{White}},
  \bibinfo{journal}{Phys. Rev. A} \textbf{\bibinfo{volume}{64}},
  \bibinfo{pages}{052312} (\bibinfo{year}{2001}).

\end{thebibliography}
\end{document}